\documentclass[twocolumn,11pt]{article}

\usepackage{epsfig}

\title {Implementation of the Quantum Fourier Transform}
\author{Yaakov S. Weinstein$^*$, Seth Lloyd$^{*\sharp}$, David G. Cory$\dag$ \\
\\
$^*$ \small{d'Arbeloff Laboratory for Information Systems and Technology} \\
\small{Department of Mechanical Engineering, M.I.T., Cambridge, MA 02139}
\\
$\dag$\small{Department of Nuclear Engineering, M.I.T Cambridge, MA 02139}
\\
$\sharp$\small{Author to whom correspondence should be addressed}
\\}
  
\begin{document}

\twocolumn[

\maketitle

\begin{abstract}

The quantum Fourier transform has been implemented on a three bit nuclear magnetic resonance (NMR) quantum computer, providing a first step towards the realization of Shor's factoring and other quantum algorithms. Implementation of the QFT is presented with fidelity measures, and state tomography. Experimentally realizing the QFT is a clear demonstration of NMR's ability to control quantum systems. 
\\
\\
PACS numbers 03.67.-a, 03.67.Lx, 02.70.-c, 89.70.+c
\\
\\
\end{abstract}]

Quantum computers are devices that process information in a way that preserves quantum coherence. Unlike a classical bit, a quantum bit, or `qubit,' can be in a superposition of $0$ and $1$ at once. This nonclassical feature of quantum information allows quantum computers to perform some computations faster than classical computers. For example, quantum computers, if constructed, could factor large numbers more rapidly \cite{Shor}, search data basis more quickly \cite{Lov}, and simulate quantum systems more efficiently \cite{Lloyd} than is possible using current classical algorithms \cite{Ben} \cite{Fey} \cite{Deu}  \cite{Steane} \cite{DiV} \cite{ref} \cite{Dan2} \cite{ch}.

A key subroutine of algorithms for factoring and simulation \cite{Dan} \cite{Zalka} \cite{Wies} is the quantum Fourier transform (QFT) \cite{cop} \cite{Ekert} \cite{Josza}. In essence the QFT takes a `position' state $|x\rangle$ to the corresponding `momentum' state $|p\rangle$ and is defined as follows:
\begin{equation}
QFT_q|x\rangle \rightarrow  \frac{1}{\sqrt{q}} \sum^{q-1}_{p=0} e^{2 \pi iap/q}|p\rangle.
\end{equation}
Where $q$ is the dimension of the systems Hilbert space.

In general the $QFT_q$ transforms the input amplitudes as,
\begin{equation}
QFT_q\sum_x f(x) |x\rangle \rightarrow  \sum_p \tilde{f}(p)|p\rangle.
\end{equation}
Where the coefficients $\tilde{f}(p)$ are 
\begin{equation}
\tilde{f}(p) = \frac{1}{\sqrt{q}} \sum_a e^{2 \pi iap/q} f(x).
\end{equation}
For example, the two qubit QFT corresponds to the unitary operator,
\begin{eqnarray}
QFT_4 &=&
\frac{1}{2}
\left(
\begin{array}{cccc}
1 & 1 & 1 & 1 \\
1 & i & -1 & -i \\
1 & -1 & 1 & -1 \\
1 & -i & -1 & i \\
\end{array}
\right) 
.
\end{eqnarray}
This operator shows the QFT separating the input states by 0 degrees in the first row and column, and then by 90 degrees, 180 degrees and 270 degrees, multiples of $\frac{\pi}{2}$. 

Equation (4) shows that the QFT has  effects similar to that of the classical Fourier transform. In particular, if $f(a)$ is periodic with period $r$, then $\tilde{f}(c)$ will exhibit a spike at $c=r$. This is the key to Shor's algorithm which allows a quantum computer to factor very large numbers in polynomial time. The classical Fourier transform reveals the periodicity in functions, the QFT reveals periodicity of wavefunctions.  

As formulated by Coppersmith, the QFT can be constructed from two basic unitary operations, $A_j$, operating on the {\it j}th qubit 
\begin{eqnarray}
A_j &=& 
\frac{1}{\sqrt{2}}
\left( 
\begin{array}{cc}
1 & 1 \\
1 & -1 \\
\end{array}
\right)
\end{eqnarray}
and $B_{jk}$ operating on the  {\it j}th and {\it k}th qubits
\begin{eqnarray}
B_{jk} &=& 
\left( 
\begin{array}{cccc}
1 & 0 & 0 & 0 \\
0 & 1 & 0 & 0 \\
0 & 0 & 1 & 0 \\
0 & 0 & 0 & e^{{i\theta_{jk}}}
\end{array}
\right) 
,
\end{eqnarray}  
where $\theta_{jk} = {\pi}/{2^{k-j}}$.

To implement the QFT, these gates,
\begin{equation}
B_{j,j+1}B_{j,j+2}...B_{j,L-1}A_j  
\end{equation}
are implemented on the lead bit, $j=L-1$. Repeating the above sequence of gates to all $L$ bits as $j$ is indexed from $L-1$ to $0$ will complete the QFT. This sequence of quantum logic gates can be realized NMR. The idea of using nuclear spins as the basic unit of a quantum computer was proposed by Lloyd \cite{Ll}, and detailed schemes for using NMR as a method of quantum computing were proposed by Cory {\it et al} \cite{Cory} and Gershenfeld and Chuang \cite{Gersh}. In NMR a series of radio frequency pulses are used to control the excess magnetization of an ensemble of quantum states. NMR experiments are easily visualized by picturing  the excess magnetization as a vector pointing in some direction and the pulses as rotations about the various axes. In addition, a bilinear coupling term in the Hamiltonian allows for quantum superposition. 

The Hamiltonian of a three spin (qubit) NMR sample with $J$-coupling is 
\begin{eqnarray}
H &=& 
\begin{array}{c}
\omega_1I^z_1+\omega_2I^z_2+\omega_3I^z_3+ \\
2\pi( J_{1,2}I^z_1 I^z_2+ J_{1,3}I^z_1 I^z_3+ J_{2,3}I^z_2 I^z_3)
\end{array}
\end{eqnarray}
where $I_i=\sigma_i/2$. The three bit QFT was implemented via NMR using the three carbon-13 spins of an alanine sample. The resonant frequency of carbon-13 at 9.4 Tesla is approximately 100.617MHz. The carbonyl was labeled spin 1, $C_\alpha$ was labeled spin 2, and $C_\beta$ spin 3. The chemical shift of the three alanine carbons are 12587Hz, 0Hz, and -3435Hz respectively. Coupling constants between the three spins are $J_{12}$ = 54Hz, $J_{23}$ = 35Hz, and $J_{13}$ = 1.2Hz. Relaxation time $T_1$ for alanine is approximately 1.56s while $T_2$ is about 420ms.

The $A_j$ matrix described above can be broken up into idempotents $E_+ - E_- + \sigma_x(E_+ + E_-)$. The pulse sequence of the $A_j$ gate can now be determined using the geometric algebra formalism \cite{shy}, 

\begin{equation}
A_j = \left(\frac{\pi}{2}\right)_y^j - \left(\pi\right)_{x}^j.
\end{equation}   
This pulse program reads: apply a pulse along the $y$-axis that rotates spin $j$ 90 degrees, apply a pulse along the $x$-axis that rotates $j$ 180 degrees. Magnetization on the $z$-axis would be rotated to the positive $x$-axis. Since this experiment starts with the spins at thermal equilibrium (pointing along the $z$-axis) the above sequence for the $A_j$ gate can be replaced by the simpler $\frac{\pi}{2}$ pulse along the positive $y$-axis.
 
The $B_{jk}$ gate, which can be constructed using the coupling between qubits, In terms of idempotents reduces to $1 - {E_-}^1{E_-}^2 + e^{i\theta}{E_-}^1{E_-}^2$. Again using geometric algebra this yields the following pulse sequence:

\begin{eqnarray}
B_{jk} &=&
\begin{array}{c}
\left(\pi\right)_{\phi}^{j} - \left(\frac{\theta}{2\pi J_{jk}}\right) - \left(\pi\right)_{\phi}^{j}\\
\\
\left(\frac{\pi}{2}\right)_y^{j,k} -\left(\frac{\theta}{2}\right)_x^{j,k} -\left(\frac{\pi}{2}\right)_{-y}^{j,k}.  
\end{array}
\end{eqnarray}

The notation $\theta/2\pi J_{jk}$ represents an interval of spin evolution under the coupling Hamiltonian. The final three pulses effectively perform a rotation around the $z$-axis. These pulses are not necessary, however, since the same effect may be achieved by rotating the prior pulses of the experiment.

The complete pulse program is the combination of $A_j$ and $B_{jk}$ gates described above. In this implementation, the necessity of performing a swap gate has been removed by reordering the bits at the appropriate interval.
 
The complete pulse program is, 

\begin{eqnarray}
QFT_3 &=&
\begin{array}{c}
\left(\frac{\pi}{2}\right)_{-sin(\frac{3\pi}{8})x+cos(\frac{3\pi}{8})y}^{1} - \left(\pi\right)_{x}^{2}\\
\\ 
\left(\frac{1}{8 J_{12}}\right) - \left(\pi\right)_{x}^{3} - \left(\frac{1}{8 J_{12}}\right) - \left(\pi\right)_{-x}^{2}\\
\\
\left(\frac{\pi}{2}\right)_{\frac{-x+y}{\sqrt{2}}}^{2} -
\left(\frac{1}{16 J_{13}}\right) -\left(\pi\right)_{x}^{2} - \\
\\
\left(\frac{1}{16 J_{13}}\right) - \left(\pi\right)_{-x}^{2} - \left(\frac{1}{8 J_{23}}\right) - \left(\pi\right)_{x}^{1} - \\
\\
\left(\frac{1}{8 J_{23}}\right) - \left(\pi\right)_{x}^{2} - \left(\pi\right)_{-x}^{1} - \left(\pi\right)_{-x}^{2} - \\
\\
\left(\pi\right)_{-x}^{3} - \left(\pi\right)_{x}^{2} -\left(\frac{\pi}{2}\right)_{y}^{3} - \left(\pi\right)_{-x}^{2}.
\end{array}
\end{eqnarray}
This sequence includes a number of $\left(\pi\right)$ pulses to refocus couplings during the intervals they should be inactive.

The pulse sequence takes advantage of knowledge of the starting state of the system at the beginning and end of the program by replacing Hadamard transforms with $\frac{\pi}{2}$ pulses. In the middle of the sequence the full Hadamard was indeed used. 

\begin{figure}
\epsfig{file=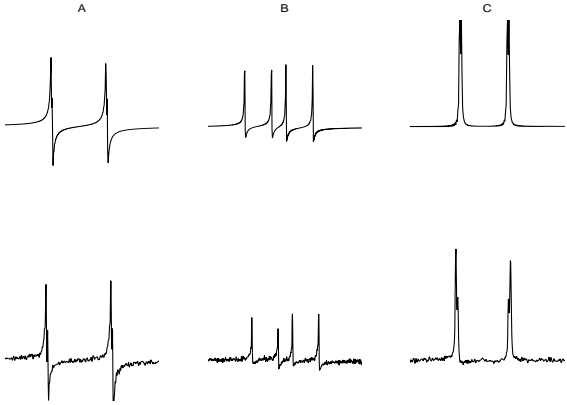, height=5.8cm, width=8cm}
\caption {The three carbon-13 atoms of alanine after performance of the QFT. The top spectra are theoretical while the bottom are experimental. Peaks in NMR spectra show the difference in energy level of single spin flips. Each spin has four peaks since its energy level is dependent on whether the other two spins are up (along the magnet) or down. This shows up clearly in spin $B$ which has resolved $J$-couples to both of the other spins. The $J$-coupling between the $A$ and $C$ spins is very small and, therefore, the four peaks are not totally resolved. These peaks tell the magnitude of only some of the terms of the density matrix.}
\end{figure}

Figure 1 shows selected theoretical and experimental spectra following the quantum Fourier transform of the state $I^{1}_{z}+I^{2}_{z}+I^{3}_{z}$ on the three qubit NMR quantum computer. 

The fidelity of the QFT calculated using the measure
\begin{equation}
F=\frac{1}{2}+\frac{1}{2} \frac{Tr(\rho_{theory}\rho_{exp})}{\sqrt{Tr(\rho_{theory}^{2})}\sqrt{Tr(\rho_{exp}^{2})}}
\end{equation}
is 87\%. Here $\rho$ is the density matrix minus the part that is proportional to the identity (in NMR, this is called the `reduced' density matrix; it should not be confused with the reduced density matrix got by partially tracing the density matrix for a composite quantum system over some of its subsystems). This measure reflects both imperfections in the applied pulses and delays, as well as decoherence. To a first approximation, decoherence during the course of the QFT attenuates the entire density matrix. This is shown in figure 2. Therefore, we can approximately separate the errors caused by experimental imperfections by renormalizing $\rho_{exp}$ to its attenuated average. Using this the fidelity of the operations themselves is above 98\% over the 6 gates in (11).

The fidelity of 87\% corresponds to an error rate of 97.7\% over the six gates which, while high, does not attain the error rate of $10^{-4}$ required for robust quantum computation \cite{Knill}. These errors arise primarily from spatial inhomogeneities in the radio frequency fields which we believe can be improved. 
 
In conclusion, using NMR, the QFT has been implemented on a three bit quantum system and the fidelity with which we can transform an initially diagonal state has been measured. Although the fidelity does not reach that required for fault tolerant computing, it is easily high enough to permit studies on small quantum systems including quantum simulations. A particularly straightforward use of the QFT is in quantum chaos: as Balazs and Voros \cite{B&V} pointed out, a simple version of the quantum baker's map can be performed by QFTs and Schack \cite{Sch} has shown how such a quantum map might be realized on a quantum computer \cite {Zurek}. 

The authors thank S. S. Somaroo and C. H. Tseng for helpful discussions. This work was supported by DARPA.
 
\begin{figure}
\epsfig{file=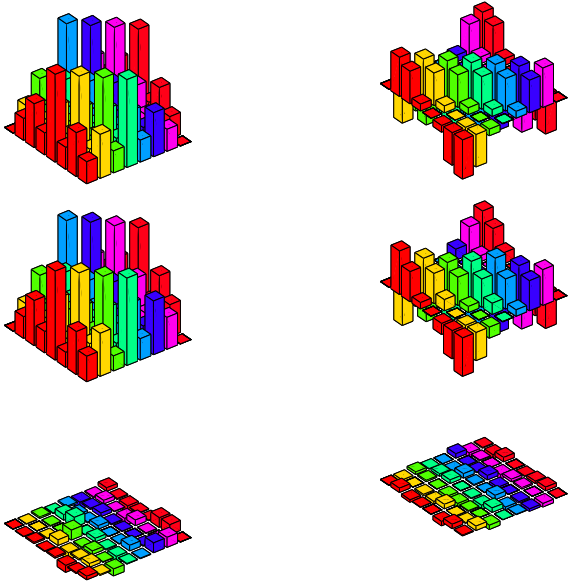, height=8cm, width=8cm}
\caption {Theoretical and experimental results of the final density matrix after implementation of the QFT on a thermal state. The left column shows (from top to bottom) the theoretical, experimental and difference of the real components of the three spin density matrix. The right column shows the same for the imaginary terms. To read all the terms of the density matrix it is necessary to rotate them into single spin single quantum terms. The diagonal of the density matrix can be seen running horizontally from the left corner to the right corner, the magnitude of all terms on the diagonal being zero. The states are labeled from $|000\rangle$ at the left and count up to $|111\rangle$ at the back and front corners.}
\end{figure}

\end{document}